# Enhanced superconductivity and superconductor to insulator transition in nano-crystalline molybdenum thin films

Shilpam Sharma[1], E. P. Amaladass[1], Neha Sharma[2], V. Harimohan[1], S. Amirthapandian[3], Awadhesh Mani[1*]

[1]Condensed Matter Physics Division, Materials Science Group, Indira Gandhi Centre for Atomic Research, Kalpakkam 603102, India
[2]Surface & Nanoscience Division, Materials Science Group, Indira Gandhi Centre for Atomic Research, Kalpakkam 603102, India
[3]Materials Physics Division, Materials Science Group, Indira Gandhi Centre for Atomic Research, Kalpakkam 603102, India

Disorder driven superconductor to insulator transition via intermediate metallic regime is reported in nano-crystalline thin films of molybdenum. The nano-structured thin films have been deposited at room temperature using DC magnetron sputtering at different argon pressures. The grain size has been tuned using deposition pressure as the sole control parameter. A variation of particle sizes, room temperature resistivity and superconducting transition has been studied as a function of deposition pressure. The nano-crystalline molybdenum thin films are found to have large carrier concentration but very low mobility and electronic mean free path. Hall and conductivity measurements have been used to understand the effect of disorder on the carrier density and mobilities. Ioffe-Regel parameter is shown to correlate with the continuous metal-insulator transition in our samples.

**\*Corresponding Author:** Awadhesh Mani
Telephone: +91-44-27480500 + (Extn.) 22356
Fax: +914427480081
Email: mani@igcar.gov.in

## 1. Introduction

Even though an old area of research in material science, the studies on the effect of disorder on normal and superconducting states of metallic superconductors still remains an open problem that is enjoying a renewed interest recently [1-11]. The disorder and superconductivity (SC) have contrasting effects on the electrical properties of material: while the former causes increase in the resistivity by localizing electron wave-functions, latter gives rise to zero resistivity on account of formation of long range ordered condensate in a macroscopic quantum state. Based on Bardeen-Cooper-Schrieffer (BCS) theory of *s*-wave superconductors, Anderson postulated that superconductivity in a metallic system can't be destroyed by weak or moderate disorder but grains ceases to be superconducting below a critical size ($D_C$), at which the superconducting gap equals the energy level spacing due to energy band discretization (Kubo gap) [12]. It was however, later observed that amorphous or granular thin film of many s-wave superconductors show a superconductor to insulator transition (SIT) initiated by a change of control parameter such as an increase in disorder [5, 13-15] or application of parallel or perpendicular magnetic field [16-18].

SIT, occurring in amorphous or granular thin films is an example of continuous quantum phase transition that transforms a coherent superconducting state in to a localized insulating ground state across the phase boundary [19]. The transition from superconducting to insulating state, tuned by change of some control parameter, could be a continuous transition such as seen in ultrathin films of amorphous Bi [20] or there could appear an intermediate, quantum corrected localized metallic regime (M) like the S-M-I transition reported in $Nb_xSi_{1-x}$ alloy films [14, 21, 22]. In the metallic regime of the split S-M-I transition, the conductivity extrapolates to a finite value at T = 0 but the resistance remains much lower than that of the normal state [22]. The theoretical understanding of the SIT is based on mainly two models: the Bosonic model presumes that Cooper pairs exist in insulating state and transition occurs due to Bose-Einstein condensation [19, 23] whereas, the Fermionic model considers that in the dirty system at zero temperature, the electrons do not pair and normal Fermionic state electrons undergo Anderson localization [23].

In addition to occurrence of SIT, superconducting properties such as transition temperature ($T_C$), superconducting gap and upper critical field ($H_{C2}$) show strong dependence on the degree of disorder or size of the superconducting grains embedded in the amorphous material. Increasing disorder by reducing particle size is known to increase or decrease $T_C$ depending upon the coupling strength and the smearing of density of states at the Fermi level [24, 25]. Weakly coupled type I superconductors, such as Al, Sn, In, W and Re, show an increase in their $T_C$ in the disordered state [24, 26, 27]. Type II superconductor with intermediate coupling like Nb shows a decrease in the superconducting gap and $T_C$ with reduction in particle size [28]. For strongly coupled Pb, while many reports on thin films exhibit no size dependent change in $T_C$ [27], there are reports of reduction in $T_C$ of Pb powder and thin films with grains below a critical size mainly due to discretization of electronic energy levels [29, 30].

There are few reports of enhancement of $T_C$ in thin amorphous films of Mo deposited at liquid helium temperature [31] but here we report disorder induced SIT via intermediary metallic phase in molybdenum thin films, sputter deposited at room temperature under varying argon pressure. In addition to transition into insulating state, our nano-crystalline thin films show enhancement in $T_C$ as reported in amorphous Mo films [24]. Detailed structural characterization and analysis using high resolution transmission electron microscopy (HRTEM) have been performed to understand the effect of deposition pressure on the grain size and normal state resistivity. Effective disorder in these films has been characterized by the Ioffe-Regel parameter ($k_F l$) [32] ($k_F$ is the Fermi wave vector and $l_e$ is the electron mean free path). A striking correlation in deposition pressure, disorder and particle size with superconducting and normal state properties of the Mo films has been brought about from these analyses.

## 2. Experimental Details

Nano-crystalline thin films of Mo were deposited on the cleaned glass substrates using a home built magnetron deposition system. Sputter deposition was carried out from a 99.97% pure Mo target using a custom built 1" magnetron cathode. Deposition chamber can reach up to a base pressure of ~$1\times10^{-7}$ mbar and the oxygen partial pressure measured using residual gas analyzer was found to be less than $10^{-9}$ mbar. The average particle size in the Mo films was reproducibly controlled by varying Ar pressure in 13 μbar to 3 μbar range. Films were deposited with direct current fixed at 40 mA and voltage varying between 270 V and 325 V. Thickness measurements were performed using Dektak profilometer, Rutherford back scattering and X-ray reflectivity (XRR) measurements. As the films were found to be X-ray amorphous, phase and microstructure of the nano-crystalline films was characterized using transmission electron microscopy (TEM) on mechanically scratched films transferred onto carbon grid. High resolution TEM measurements were performed using LIBRA 200FE HRTEM operated at 200 keV, equipped with an in-column energy filter and Schottky field emission gun source. Electron diffraction patterns (SAED) and electron energy loss spectrums (EELS) were collected at different regions of the samples to confirm purity of Mo phase in the films. In-field transport measurements were performed using M/S cryogenic Inc. U.K. make cryogen free, 15 T magneto-transport cryostat capable of reaching temperatures as low as 1.8 K from 300 K. Hall and resistivity measurements were performed in Hall bar and Van der Pauw geometry respectively [33, 34].

## 3. Results and Discussion

All the nano-crystalline Mo films were deposited on glass substrates at different Ar pressures varying from 3 μbar to 13 μbar while maintaining same DC current of 40 mA. The thickness of all the films was found to be roughly 56 nm as measured using Rutherford backscattering, Dektak profilometer and XRR measurements.

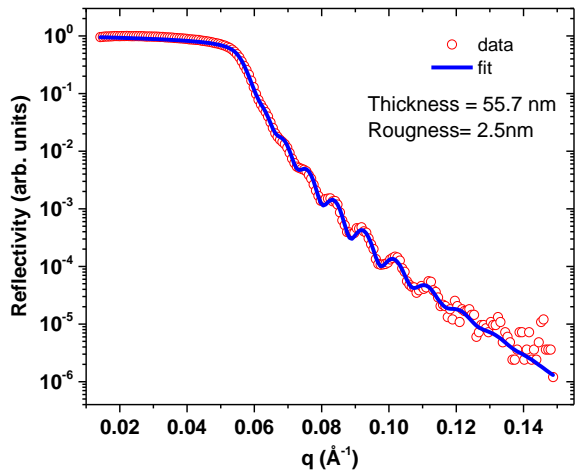

**Figure 1:** X-ray reflectivity as a function of scattering vector. Thickness of film was estimated by fitting the XRR data. A thickness of ~56 nm with a low surface roughness of ~2.5 nm has been observed.

To estimate film thickness and roughness, XRR data was fitted using Parratt32 routine [35]. A representative XRR data along with fit is presented in figure 1. The surface roughness of the films was found to be ~2.5 nm. These Mo thin films were characterized for phase and microstructure using HRTEM. Figure 2(a)-(d) show the HRTEM images of the films deposited at different Ar pressures 3.0, 4.5, 7.0 and 13.0 μbar respectively. It can be clearly seen that the film deposited at 3 μbar has large particles with sharp particle boundaries and as the deposition pressure increases, particle sizes reduces. In the HRTEM image (c.f. Figure 2(e)) of sample deposited at 3 μbar Ar pressure, two grains can be identified with crystalline order ranging up to 5-6 nm with inter-granular region of roughly 2 nm. With increase in deposition pressure, inter-granular region comprising of amorphous Mo with no long range order but some degree of short range order is seen to increase.

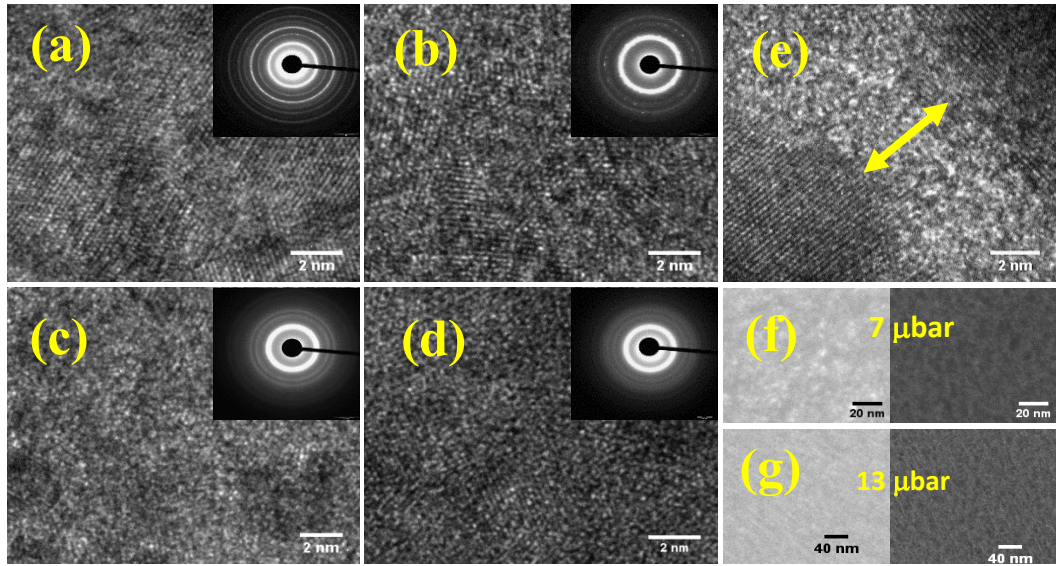

**Figure 2 (a-d) (Color Online):** HRTEM images of Mo thin films deposited at Ar pressure 3.0, 3.5, 4.5, 5.5 μbar respectively. Grain size reduces and inter-granular region with short range order increases with increase in deposition pressure. SAED patterns of samples deposited at different pressures show a change of polycrystalline ring pattern to diffused ring pattern typical of amorphous materials. **(e):** shows intergrain amorphous region of around 2 nm between grains having long range order up to 5-6 nm. **(f)-(g):** bright field and dark field images for samples deposited at 7 and 13 μbar pressure shows that the particle density reduces with increase in deposition pressure.

Figure 2(a)-(d) also show corresponding SAED patterns collected from the samples. It can be observed from figure 2(a) that the film deposited at 3 μbar is polycrystalline in nature with well-defined Debye-Scherrer ring pattern. The ring pattern was analyzed using "process diffraction" software [36]. All the d-spacing in SAED pattern were indexed to BCC Mo with a lattice parameter ~2.9404 Å, smaller than the reported value of 3.1472 Å [37] thus indicating a compressive stress in the film. Similar changes in lattice parameters of nano-crystalline films of Nb and gold have already been reported in literature [28, 38, 39]. Other than BCC Mo d-spacings, the d-spacing corresponding to impurity phases like molybdenum oxide etc. have not been observed in the diffraction pattern. With increase in the deposition pressure, Debye-Scherrer rings

in the SAED patterns (c.f. figure 2b-d) can't be discerned and the patterns resemble diffraction pattern of amorphous material with diffused rings. For sample deposited at 4.5 µbar pressure (figure 2(b)), a few crystalline spots can be observed on the amorphous electron diffraction pattern. Figure 2(f) and 2(g) show dark field and bright field TEM micrographs for 7 µbar and 13 µbar deposited samples. Dark contrast due to diffracting grains can be observed. It is seen that the number density of the particles is decreasing with increase in deposition pressure.

The particle size distribution of these films has been estimated by analyzing dark field TEM micrographs using ImageJ software [40]. The average particle size has been estimated by fitting log-normal distribution to the size distribution histograms of roughly 100-150 different particles.

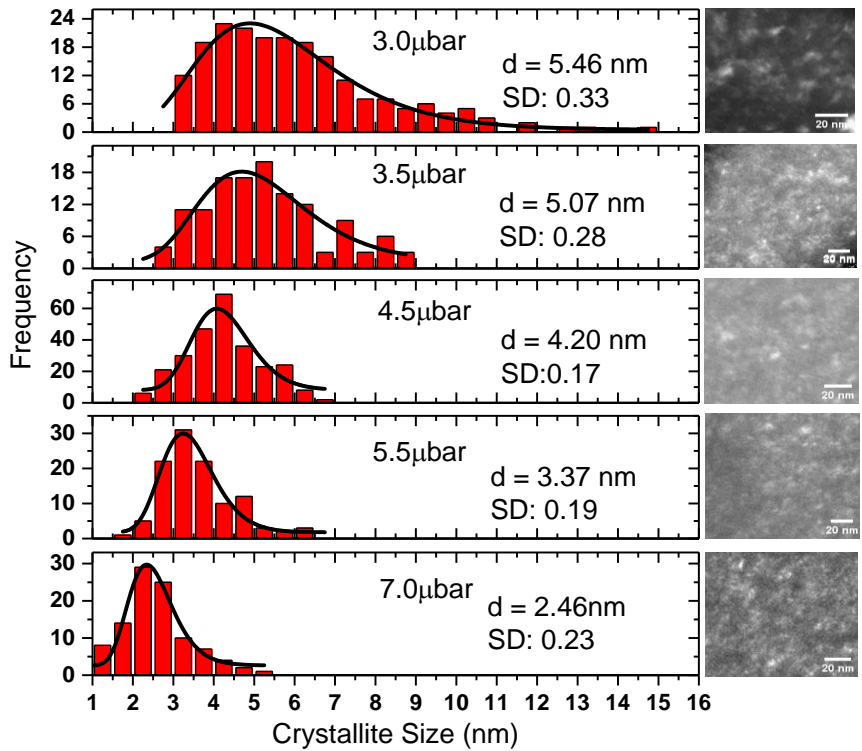

**Figure 3 (Color Online):** Particle size distribution in Mo thin films deposited at different Ar pressures from 3.0 to 7.0 µbar. A decrease in particle size is observed with increase in deposition pressure. Figure also presents dark field TEM micrographs of the corresponding thin films used to deduce particle sizes.

The particle size distribution along with a representative dark field TEM micrograph is presented in figure 3. The average particle size decreases from 5.5 nm to 2.5 nm with increase in deposition pressure (P) from 3 μbar to 7 μbar. With a further increase of pressure from 7 μbar to 13 μbar, particle size does not decrease below ~2.5 nm, nonetheless, the average particle density decreases. The variation of average particle size with deposition pressure is presented in figure 4. It can be seen that particle size decreases rapidly as the pressure is increased from 3 μbar to 7 μbar and beyond P > 7 μbar it remains nearly constant at ~2.5 nm. A smaller than reported value of the lattice parameter of our films can thus be due to pressure from grain boundary on these small crystallites. Nano-structured Nb thin films deposited at same Ar pressure were reported to have ~60 nm particles and the smallest particle of ~5 nm for 13 μbar deposition pressure [41].

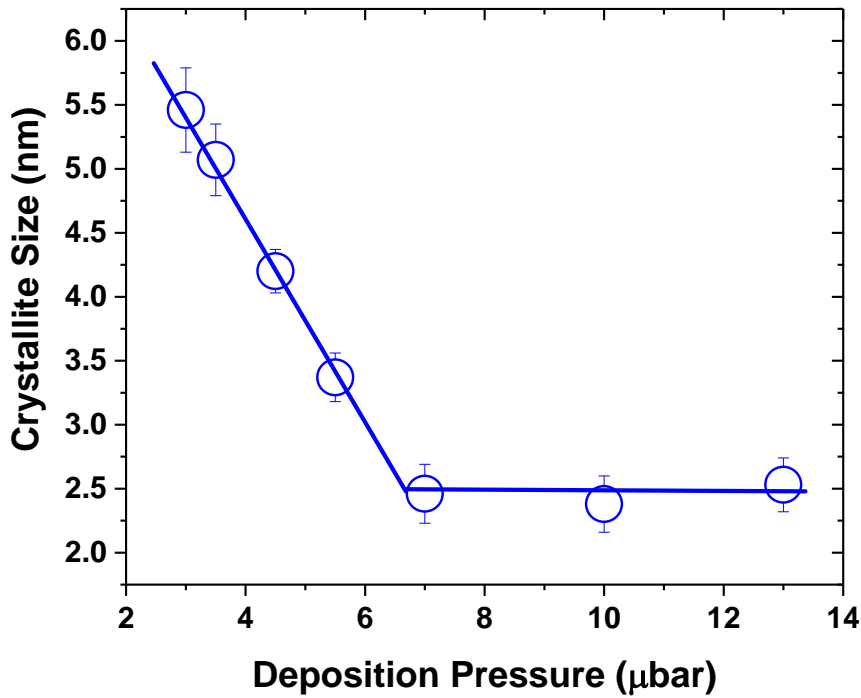

**Figure 4 (Color Online):** Variation of average particle size with deposition pressure. A quick fall in particle size is observed when deposition pressure was increased from 3 μbar to 7 μbar. Above 7 μbar the particle size remains constant at ~2.5 nm. Blue lines are for guidance.

Figure 5 shows the temperature variation of resistivity of nano-crystalline Mo thin films. Superconducting transition is found to occur in samples with particle sizes up to 3.4 nm (i.e. films deposited up to P = 5.5 µbar). Resistive transitions to the superconducting state are shown in figure 6a. $T_C$ is well defined and much higher than bulk $T_C$ ~ 915 mK, thus providing evidence of size effects due to nano-crystallinity of our Mo thin films. Variation of $T_C$ with pressure along with respective grain size is shown in figure 6b. A dome shaped variation of $T_C$ is seen with maximum at ~5.5 K occurring for the film with 4.2 nm particle size deposited at 4.5 µbar pressure. Similar enhancement and subsequent suppression of superconducting $T_C$ as a function of a tuning parameter has already been reported in granular Al [42] and Pb thin films [43].

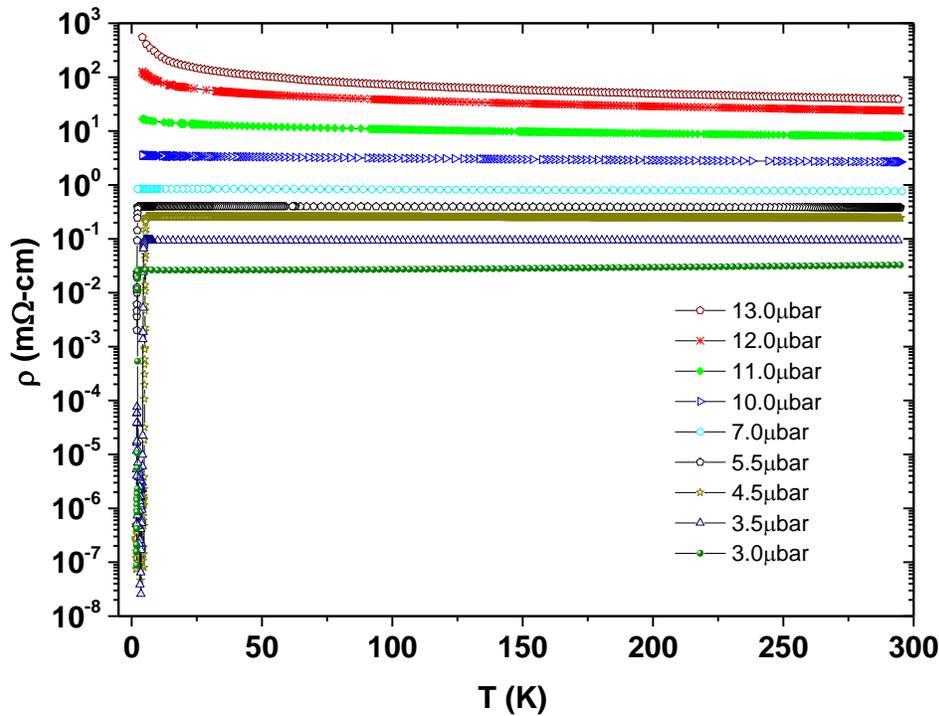

**Figure 5 (Color Online)**: Temperature variation of normalized resistance ($R_T/R_{299\ K}$). Sample deposited at 3.0 µbar pressure shows positive TCR.

For the amorphous thin films of weakly coupled Mo, $T_C$ gets enhanced to as high as 6.7 to 8 K [24, 25] and there are reports that $T_C$ increases marginally with increase in stress [44, 45] and is correlated with sheet resistance [31]. The non-monotonic dome shaped variation of $T_C$ with deposition pressure or particle size in our Mo thin films has resemblance to SC phase diagram of high $T_C$ superconductors [46, 47]. The $T_C$ is basically governed by the amplitude (SC energy gap) and phase of the complex SC order parameter. In addition, the rigidity of the global phase coherence depends upon superfluid stiffness which for strongly coupled grains exceed the SC energy gap and the $T_C$ in this case is amplitude driven. If the coupling between the grains is weak then the superfluid stiffness gets suppressed and the transition to normal state occurs due to phase fluctuations between neighboring SC grains, giving rise to dissipative current flow. The dome shape of the phase diagram can thus be due to a crossover from $T_C$ determined from phase fluctuations induced decoherence of weakly coupled SC nano-grains in disordered matrix to the $T_C$ due to amplitude suppression in large grains forming a strongly coupled Josephson array [42, 43].

There are two propositions to explain the enhancement in the $T_C$ of nano-crystalline weakly coupled SC: first, the $T_C$ can be enhanced due to the surface phonon softening on the surface of nano-grains that can lead to an increase in electron-phonon coupling strength with increase in the surface to volume ratio of the grains [30] or secondly, there could be shell effect due to discretization of the continuous bulk density of states in to energy levels in the nano-structured grains. The size induced fluctuations in the number of discrete levels around the Fermi energy can increase the spectral density thereby increasing pairing and SC gap [48, 49]. Since our Mo films are granular with grains less than 5-6 nm in size, the $T_C$ enhancement could be due to finite size effects. The decrease in the $T_C$ for the films deposited at higher pressure (P > 4.5 µbar) could be due to increase in disorder leading to reduction in phase stiffness and decoherence between the SC grains. On the other hand, for the films deposited at lower pressure (P < 4.5 µbar), even though the intergranular coupling could be more strong but since the crystallinity increases with decreasing pressure, the $T_C$ starts decreasing from its maximum value in disordered state to finally approaching towards its bulk value in crystalline Mo.

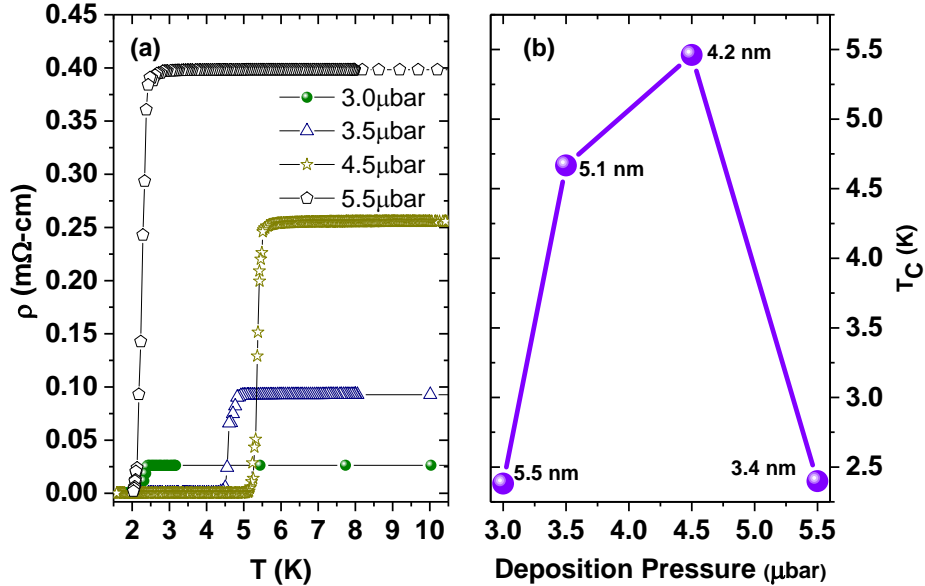

**Figure 6(a) (Color Online)**: The resistive transitions to superconducting state as a function of temperature for films deposited at different pressures. The R(T) plots do not show reentrant behavior and the $T_C$ is well defined and higher than bulk $T_C \sim 915$ mK. **Figure 6(b):** Variation of $T_C$ with deposition pressure and particle size. $T_C$ increases up to 5.5 K for sample deposited at 4.5 μbar pressure and reduces on further increase in deposition pressure. Superconducting transition was not observed down to 2 K for sample deposited at 7.0 μbar.

As evident from figure 5, upon increasing the deposition pressure, a change in the temperature coefficient of resistance (TCR) has been observed. Film with average particle size ~5.5 nm (deposited at P = 3 μbar) shows positive TCR with a residual resistivity ratio (RRR = $R_{299 K}/R_{6 K}$) of ~1.2, whereas a RRR ~1 has been observed for the sample deposited at P = 3.5 μbar with little temperature dependence of normal state resistivity up to superconducting transition. Films deposited at P ≥ 4.5 μbar showed negative TCR with RRR as low as 0.07. A plot of RRR for all the films with respect to deposition pressure is presented in figure 7. RRR decreases monotonically with increase in deposition pressure but the rate of decrease of RRR becomes larger for the films deposited at P > 7 μbar. Variation of room temperature resistivity ($\rho_{299}$) of these thin films as a function of deposition pressure is also presented in figure 7.

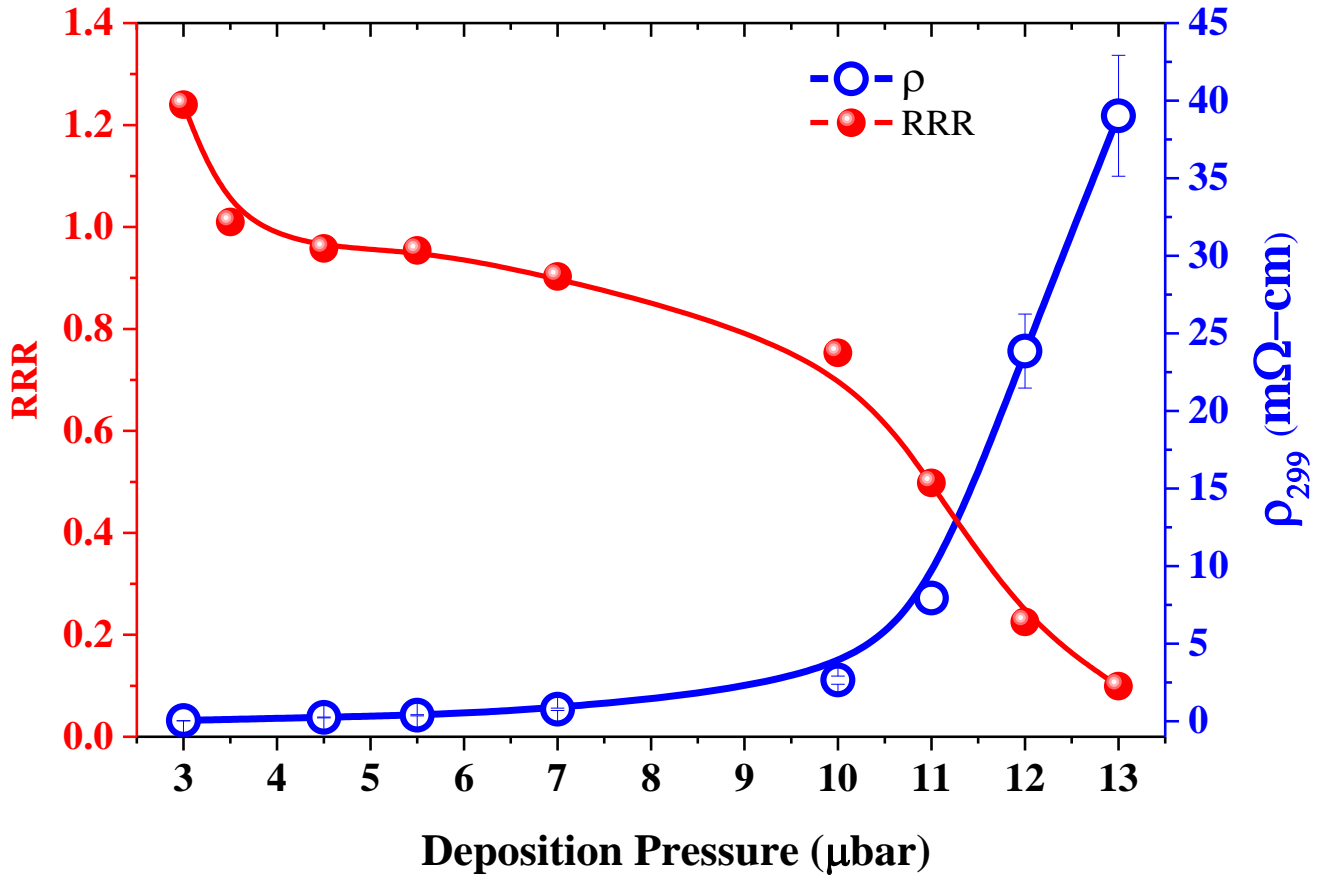

**Figure 7 (Color Online):** Variation of room temperature resistivity and RRR as a function of deposition pressure. The increase in room temperature resistivity is small up to 7 μbar and it increases faster for samples deposited at further higher pressures.

A slow increase in room temperature resistivity was observed for samples deposited up to P = 7 μbar pressure, beyond which room temperature resistivity increases rapidly. A comparison with variation of particle size as a function of deposition pressure (cf. figure 4) reveals that the resistivity increases slowly when particle size decreases rapidly from 5.5 nm to 2.5 nm but then increases rapidly in the range where particle size remains constant at 2.5 nm.

To understand a rapid increase in the resistivity of Mo films deposited at P > 7 μbar, Ioffe-Regel parameter ($k_F l$) which characterize the disorder was calculated from room temperature (299 K) Hall coefficient and resistivity. Here $k_F$ is the Fermi wave vector and $l$ is the electronic mean free path. Ioffe-Regel criterion describes transition of a metal into an insulator primarily due to increased disorder and reduced carrier density. Electrons in a metal localizes when the mean free path reduces below Fermi wave vector and the system can transition into a weak insulating phase that remains fairly conducting at any finite temperatures. In a non-interacting system with free electrons, $k_F l$ offers a distinctive measure of electronic disorder. This however is not true in a dielectric state where electron-electron (e-e) interactions cannot be ignored. Thus the $k_F l$ calculations were performed from Hall coefficient and resistivity measured at room temperature, where e-e interaction effects are not very large [50]. Ioffe-Regel parameter was calculated using the free electron formula $k_F l = \{(3\pi^2)^{\frac{2}{3}} \hbar R_H^{\frac{1}{3}}\} / [\rho e^{5/3}]$ ($\hbar$ is Planck's constant, $R_H$ is Hall coefficient, ρ is resistivity and e is the electronic charge) [32]. Hall measurements were performed on samples deposited in Hall bar geometry. A standard four probe AC technique using lock-in amplifier to record transverse resistances ($R_{xy}$) as a function of magnetic field has been used. The Hall resistance as a function of magnetic field for samples deposited at different deposition pressures is presented in figure 8. A positive slope of $R_{xy}$ (B) plots indicate that holes are the dominant carriers. It is already reported [51] that Mo is a compensated metal with holes having higher mobility ($\mu_h$ ~ $2\mu_e$ at 300 K) than electrons and thus shows positive Hall voltage. Variation of Hall coefficient as a function of deposition pressure is shown in inset (a) of figure 9. Hall coefficient increases with increase in deposition pressure. This could be due to decrease in mobility of charge carriers or decrease in the number of charge carriers taking part in conduction.

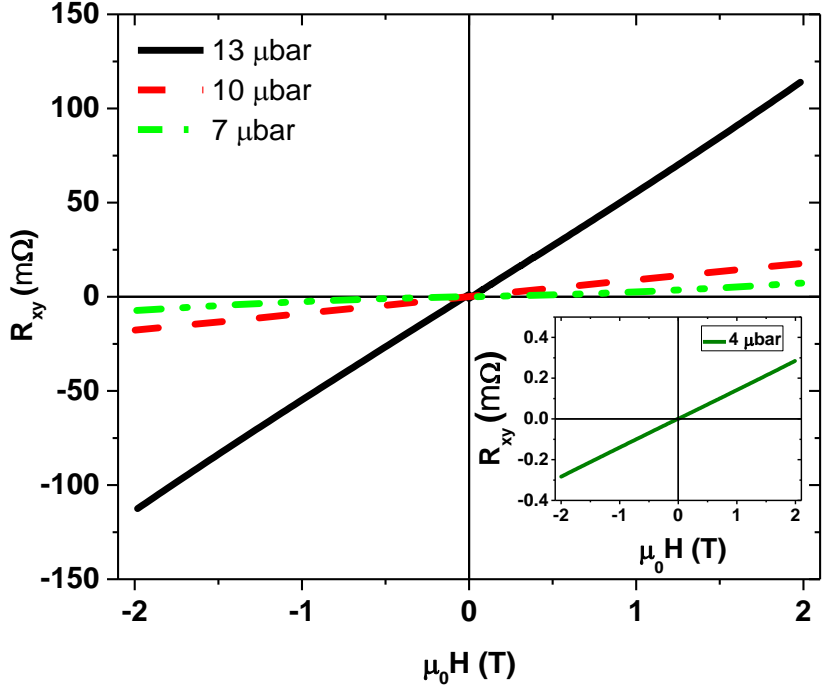

**Figure 8 (Color Online):** The Hall (transverse) resistance as a function of magnetic field for samples deposited at different deposition pressures. A positive slope of $R_{xy}$ (B) plots indicate that holes are the dominant carriers. The data is presented after subtracting the longitudinal resistance of the samples.

The value of Hall coefficient in our disordered Mo films is very close to the value reported for 30 nm thick amorphous Mo film [51, 52]. It could be observed from inset (b) of the figure 9 that while mean free path of charge carrier does not change much up to P = 7 µbar but thereafter reduces by one order of magnitude for insulating samples, the charge carriers shows a monotonous decrease by two orders of magnitude with increase in deposition pressure. This reduction in charge carrier density could be due to different carrier concentrations in crystalline and amorphous Mo or charge carrier localization in the individual but sparsely populated grains due to lack of percolation path in highly disordered samples. The charge carrier concentration varies between $10^{27}$ to $10^{29}$ m$^{-3}$ which is consistent with the reported value [52]. Even though we find a high carrier concentration in our samples, the mobility ~ $10^{-1}$ cm$^2$/V-sec. and electron mean free path ~$10^{-10} - 10^{-11}$ m are several orders of magnitude lower than that of metals or semiconductors and in particular from that of Mo metal [51, 53]. These low values of charge mobility and mean free path are commensurate with the assertion of the high degree of disorder in our samples.

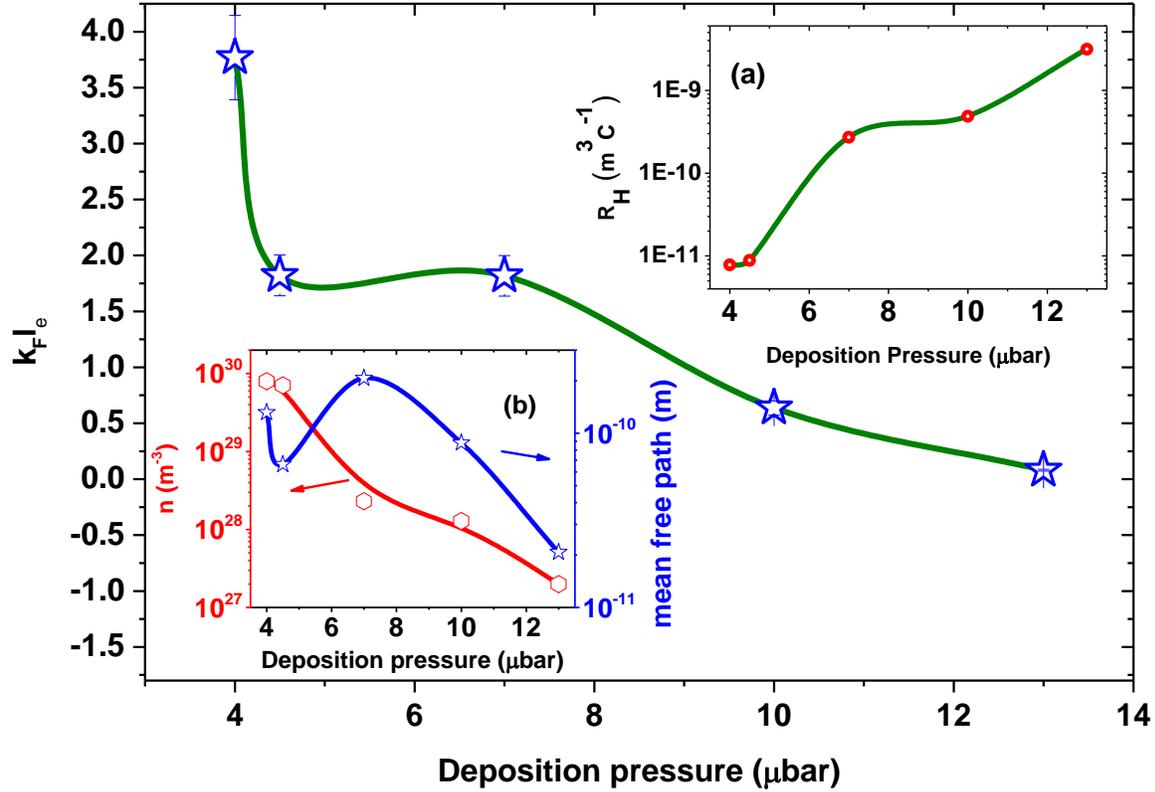

**Figure 9 (Color Online):** Ioffe-Regel parameter plotted as a function of deposition pressure. $k_F l$ shows a monotonous decrease with increase in deposition parameter. The value of $k_F l$ is >1 for samples deposited at P ≤ 7 μbar. MIT occurs for samples having $k_F l$ < 1.0 deposited at P ≥ 10 μbar **Inset (a):** Plot of $R_H$ vs. deposition pressure shows an increase in Hall coefficient with increase in pressure. **Inset (b):** variation of charge carrier density and electron mean free path as a function of deposition pressure. It is evident that the carrier density is reducing by increasing deposition pressure indicating a disorder driven localization. The mean free path is seen to be several orders of magnitude less than that in conventional metals but is found to be independent of the deposition pressure up to 7 μbar but after that reduces by one order for insulating samples.

Main panel of figure 9 shows variation of $k_F l$ as a function of deposition pressure. With increase in deposition pressure, $k_F l$ decreases monotonically from ~3.8 to 0.08. As can be seen from figure 9, the films deposited at pressures between 3-7 μbar have $k_F l$ >1, whereas, $k_F l$ <1 for films deposited at P ≥ 10 μbar. Thus all the samples deposited at pressures higher than 10 μbar become weakly insulating due to disorder induced charge carrier localization as per the Ioffe-Regel criterion.

To substantiate the existence of insulating state in samples having $k_F l < 1$, an analysis of the logarithmic derivative of conductivity ($\sigma$) as a function of temperature has been performed based on the technique introduced by Mobius et al. [54, 55]. The plot of mathematical function $w(T) = \frac{d\ln\sigma}{d\ln T}$ shows distinctively different behavior for insulating and metallic films. While the $w(T)$ plot diverges to infinity as the temperature is reduced for an insulating sample, it extrapolates to zero at zero temperature for metallic samples. The function $w(T)$ has been calculated using linear regression fit of $\ln\sigma$ vs. $\ln T$ plots. Plots of $w(T)$ as a function of $T^{1/2}$ for samples deposited at different pressures are shown in figure 10.

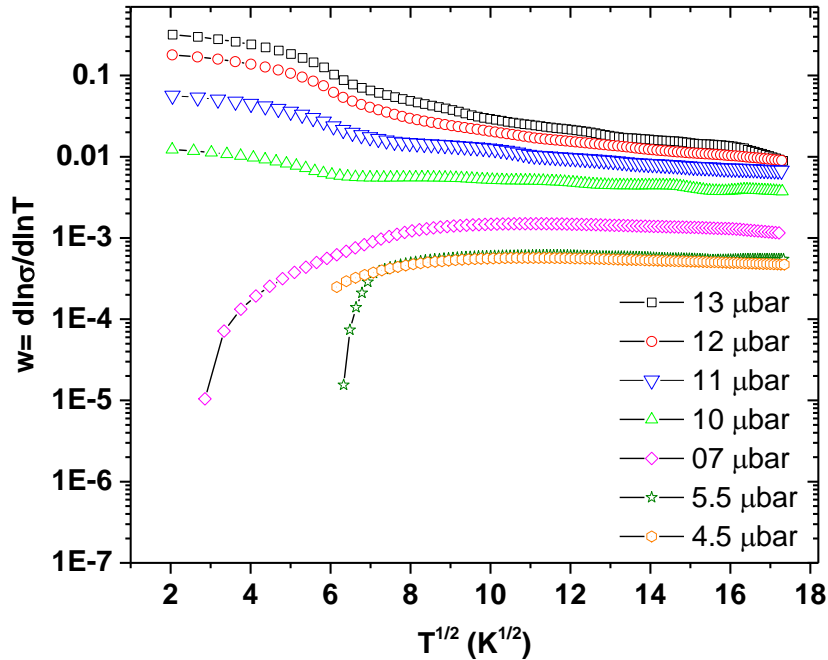

**Figure 10 (Color Online):** Plot of logarithmic derivative of conductivity as a function of temperature shows a metal-insulator transition occurring in disordered Mo films. Samples deposited at $P \leq 7$ μbar show amorphous metallic character and the samples deposited at $P \geq 10$ μbar show weakly localized insulating phase.

It can be observed that the *w(T)* values for samples deposited up to 7 μbar pressure appear to extrapolate to zero with decrease in sample temperature. This downward trend at low temperature is a manifestation of metals or bad metals. The samples deposited at P >10 μbar show an upward trend in their *w(T)* plots with reduction in temperature which gives a clear evidence of existence of localized insulating state [56] in these samples. From the plots of *w(T)* it is now reasonably clear that there is a metal to insulator transition occurring in samples deposited between 7 μbar and 10 μbar pressure. This explains the rapid rise in room temperature resistivity for samples deposited at P ≥ 10 μbar, even though particle size remains nearly the same for these samples. A similar observation of MIT in highly disordered $RuO_2$ films having very large carrier concentrations and very low mobilities has recently been reported [10].

## 4. Conclusion

In conclusion, we have shown that the nano-crystalline thin films of molybdenum exhibit superconductivity at higher transition temperature than the bulk and their $T_C$ depends upon the grain size and disorder in the films. The transition temperature initially increases with increase in the grain size due to finite size effects but then decreases possibly due to lack of coherence between neighboring grains thereby giving a dome shaped $T_C$ vs. deposition pressure phase diagram. The superconducting transition and normal state properties of these disordered thin films could be reproducibly tuned by varying Ar pressure during deposition. An increase in $T_C$ of Mo thin films was previously reported only in quench condensed amorphous Mo thin films, where alloying was required to stabilize higher $T_C$ at room temperature [24, 25]. In present study, a large increase in the $T_C$ of room-temperature-deposited Mo thin films has been observed which remains stable even with thermal cycling to low temperatures.

The films show a split superconductor to insulator transition with an intermediate metallic regime tuned by increase in the disorder due to localization of charge carriers. The carrier density of Mo films is seen to decrease with increase in the disorder. The mean free path and mobility of charge carrier has been found to be 2-3 orders of magnitude less than that of the crystalline Mo. The analysis of Ioffe-Regel parameter and conductivity data clearly indicates that the films are granular disordered systems which undergo a superconductor to intermediate metal and then metal-insulator transition.

Acknowledgements

The authors would like to thank Dr. B. Sundaravel, Materials Science Group, IGCAR for Rutherford backscattering measurements and estimation of film thickness. Authors would also like to thank Dr. Sanjay Rai, RRCAT, Indore for X-ray reflectivity measurements. The authors gratefully acknowledge UGC-DAE-CSR node at Kalpakkam for providing access to the 15 T cryogen free magneto-resistance set up and high resolution TEM imaging.

References

[1] A. Ghosal, M. Randeria, N. Trivedi, Role of Spatial Amplitude Fluctuations in Highly Disordered *s*-Wave Superconductors, Physical Review Letters, 81 (1998) 3940-3943.

[2] M. Mondal, A. Kamlapure, M. Chand, G. Saraswat, S. Kumar, J. Jesudasan, L. Benfatto, V. Tripathi, P. Raychaudhuri, Phase Fluctuations in a Strongly Disordered s-Wave NbN Superconductor Close to the Metal-Insulator Transition, Physical Review Letters, 106 (2011) 047001.

[3] M. Mondal, M. Chand, A. Kamlapure, J. Jesudasan, V. Bagwe, S. Kumar, G. Saraswat, V. Tripathi, P. Raychaudhuri, Phase Diagram and Upper Critical Field of Homogeneously Disordered Epitaxial 3-Dimensional NbN Films, J Supercond Nov Magn, 24 (2011) 341-344.

[4] A. Kamlapure, T. Das, S.C. Ganguli, J.B. Parmar, S. Bhattacharyya, P. Raychaudhuri, Emergence of nanoscale inhomogeneity in the superconducting state of a homogeneously disordered conventional superconductor, Sci. Rep., 3:2979 (2013).

[5] M. Chand, G. Saraswat, A. Kamlapure, M. Mondal, S. Kumar, J. Jesudasan, V. Bagwe, L. Benfatto, V. Tripathi, P. Raychaudhuri, Phase diagram of the strongly disordered *s*-wave superconductor NbN close to the metal-insulator transition, Physical Review B, 85 (2012) 014508.

[6] D. Kowal, Z. Ovadyahu, Scale dependent superconductor–insulator transition, Physica C: Superconductivity, 468 (2008) 322-325.

[7] D. Sherman, G. Kopnov, D. Shahar, A. Frydman, Measurement of a Superconducting Energy Gap in a Homogeneously Amorphous Insulator, Physical Review Letters, 108 (2012) 177006.

[8] Y. Dubi, Y. Meir, Y. Avishai, Nature of the superconductor-insulator transition in disordered superconductors, Nature, 449 (2007) 876-880.

[9] D. Sherman, B. Gorshunov, S. Poran, N. Trivedi, E. Farber, M. Dressel, A. Frydman, Effect of Coulomb interactions on the disorder-driven superconductor-insulator transition, Physical Review B, 89 (2014) 035149.

[10] M.S. Osofsky, C.M. Krowne, K.M. Charipar, K. Bussmann, C.N. Chervin, I.R. Pala, D.R. Rolison, Disordered $RuO_2$ exhibits two dimensional, low-mobility transport and a metal–insulator transition, Scientific Reports, 6 (2016) 21836.

[11] N.N. Kovaleva, D. Chvostova, A.V. Bagdinov, M.G. Petrova, E.I. Demikhov, F.A. Pudonin, A. Dejneka, Interplay of electron correlations and localization in disordered β-tantalum films: Evidence from dc transport and spectroscopic ellipsometry study, Applied Physics Letters, 106 (2015) 051907.

[12] P.W. Anderson, Theory of dirty superconductors, Journal of Physics and Chemistry of Solids, 11 (1959) 26-30.

[13] A.M. Goldman, N. Marković, Superconductor-Insulator Transitions in the Two-Dimensional Limit, Print edition, 51 (1998) 39.

[14] G. Hertel, D.J. Bishop, E.G. Spencer, J.M. Rowell, R.C. Dynes, Tunneling and Transport Measurements at the Metal-Insulator Transition of Amorphous Nb: Si, Physical Review Letters, 50 (1983) 743-746.

[15] R.C. Dynes, J.P. Garno, G.B. Hertel, T.P. Orlando, Tunneling Study of Superconductivity near the Metal-Insulator Transition, Physical Review Letters, 53 (1984) 2437-2440.

[16] L. Yize Stephanie, Signature of Cooper pairs in the non-superconducting phases of amorphous superconducting tantalum films, Superconductor Science and Technology, 28 (2015) 025002.

[17] G. Sambandamurthy, L.W. Engel, A. Johansson, E. Peled, D. Shahar, Experimental Evidence for a Collective Insulating State in Two-Dimensional Superconductors, Physical Review Letters, 94 (2005) 017003.

[18] T.I. Baturina, A.Y. Mironov, V.M. Vinokur, M.R. Baklanov, C. Strunk, Localized Superconductivity in the Quantum-Critical Region of the Disorder-Driven Superconductor-Insulator Transition in TiN Thin Films, Physical Review Letters, 99 (2007) 257003.

[19] M.P.A. Fisher, Quantum phase transitions in disordered two-dimensional superconductors, Physical Review Letters, 65 (1990) 923-926.

[20] D.B. Haviland, Y. Liu, A.M. Goldman, Onset of superconductivity in the two-dimensional limit, Physical Review Letters, 62 (1989) 2180-2183.

[21] Y.-H. Lin, J. Nelson, A.M. Goldman, Superconductivity of very thin films: The superconductor–insulator transition, Physica C: Superconductivity and its Applications, 514 (2015) 130-141.

[22] N. Mason, A. Kapitulnik, Dissipation Effects on the Superconductor-Insulator Transition in 2D Superconductors, Physical Review Letters, 82 (1999) 5341-5344.

[23] F.G. Vsevolod, T.D. Valery, Superconductor–insulator quantum phase transition, Physics-Uspekhi, 53 (2010) 1.

[24] J.E. Crow, M. Strongin, R.S. Thompson, O.F. Kammerer, The superconducting transition temperatures of disordered Nb, W, and Mo films, Physics Letters A, 30 (1969) 161-162.

[25] M.M. Collver, R.H. Hammond, Superconductivity in "Amorphous" Transition-Metal Alloy Films, Physical Review Letters, 30 (1973) 92-95.

[26] S. Matsuo, H. Sugiura, S. Noguchi, Superconducting transition temperature of aluminum, indium, and lead fine particles, J Low Temp Phys, 15 (1974) 481-490.

[27] B. Abeles, R.W. Cohen, G.W. Cullen, Enhancement of Superconductivity in Metal Films, Physical Review Letters, 17 (1966) 632-634.

[28] S. Bose, P. Raychaudhuri, R. Banerjee, P. Vasa, P. Ayyub, Mechanism of the Size Dependence of the Superconducting Transition of Nanostructured Nb, Physical Review Letters, 95 (2005) 147003.

[29] W.H. Li, C.C. Yang, F.C. Tsao, K.C. Lee, Quantum size effects on the superconducting parameters of zero-dimensional Pb nanoparticles, Physical Review B, 68 (2003) 184507.

[30] B. Sangita, G. Charudatta, S.P. Chockalingam, B. Rajarshi, R. Pratap, A. Pushan, Competing effects of surface phonon softening and quantum size effects on the superconducting properties of nanostructured Pb, Journal of Physics: Condensed Matter, 21 (2009) 205702.

[31] A. Hirakawa, K. Makise, T. Kawaguti, B. Shinozaki, Thickness-tuned superconductor–insulator transitions in quench-condensed Mo and MoRu films, Journal of Physics: Condensed Matter, 20 (2008) 485206.

[32] A.F. Ioffe, A.R. Regel, Non-crystalline, amorphous, and liquid electronic semiconductors, Prog. Semicond., 4 (1960) 237-291.

[33] L.J. Van der Pauw, A Method of Measuring Specific Resistivity and Hall Effect of Discs of Arbitrary Shape, Philips Research reports, 13 (1958) 1-9.

[34] L.J. Van der Pauw, A Method of Measuring The Resistivity and Hall Coefficient of Lamellae of Arbitrary Shape, Philips Technical Review, 20 (1959) 220.

[35] C. Braun, Parratt32, in, HMI Berlin, Germany, 2002.

[36] J.L. Lábár, Consistent indexing of a (set of) single crystal SAED pattern(s) with the ProcessDiffraction program, Ultramicroscopy, 103 (2005) 237-249.

[37] R.G. Ross, W. Hume-Rothery, High temperature X-ray metallography: I. A new debye-scherrer camera for use at very high temperatures II. A new parafocusing camera III. Applications to the study of chromium, hafnium, molybdenum, rhodium, ruthenium and tungsten, Journal of the Less Common Metals, 5 (1963) 258-270.

[38] R. Banerjee, E.A. Sperling, G.B. Thompson, H.L. Fraser, S. Bose, P. Ayyub, Lattice expansion in nanocrystalline niobium thin films, Applied Physics Letters, 82 (2003) 4250-4252.

[39] P.-Y. Gao, W. Kunath, H. Gleiter, K. Weiss, The structure of small penta-twinned gold particles, Scripta Metallurgica, 22 (1988) 683-686.

[40] C.A. Schneider, W.S. Rasband, K.W. Eliceiri, NIH Image to ImageJ: 25 years of image analysis, Nat Meth, 9 (2012) 671-675.

[41] B. Sangita, B. Rajarshi, G. Arda, R. Pratap, L.F. Hamish, A. Pushan, Size induced metal–insulator transition in nanostructured niobium thin films: intra-granular and inter-granular contributions, Journal of Physics: Condensed Matter, 18 (2006) 4553.

[42] U.S. Pracht, N. Bachar, L. Benfatto, G. Deutscher, E. Farber, M. Dressel, M. Scheffler, Enhanced Cooper pairing versus suppressed phase coherence shaping the superconducting dome in coupled aluminum nanograins, Physical Review B, 93 (2016) 100503.

[43] L. Merchant, J. Ostrick, R.P. Barber, R.C. Dynes, Crossover from phase fluctuation to amplitude-dominated superconductivity: A model system, Physical Review B, 63 (2001) 134508.

[44] L. Fàbrega, A. Camón, I. Fernández-Martínez, J. Sesé, M. Parra-Borderías, O. Gil, R. González-Arrabal, J.L. Costa-Krämer, F. Briones, Size and dimensionality effects in superconducting Mo thin films, Superconductor Science and Technology, 24 (2011) 075014.

[45] L. Fabrega, I. Fernandez-Martinez, M. Parra-Borderias, O. Gil, A. Camon, R. Gonzalez-Arrabal, J. Sese, J. Santiso, J.L. Costa-Kramer, F. Briones, Effects of Stress and Morphology on the Resistivity and Critical Temperature of Room-Temperature-Sputtered Mo Thin Films, Applied Superconductivity, IEEE Transactions on, 19 (2009) 3779-3785.

[46] D.M. Broun, What lies beneath the dome?, Nat Phys, 4 (2008) 170-172.

[47] D. Johnston, The puzzle of high temperature superconductivity in layered iron pnictides and chalcogenides, Advances in Physics, 59 (2010) 803-1061.

[48] J. Mayoh, A.M. García-García, Strong enhancement of bulk superconductivity by engineered nanogranularity, Physical Review B, 90 (2014) 134513.

[49] S. Bose, A.M. Garcia-Garcia, M.M. Ugeda, J.D. Urbina, C.H. Michaelis, I. Brihuega, K. Kern, Observation of shell effects in superconducting nanoparticles of Sn, Nat Mater, 9 (2010) 550-554.

[50] B.L. Altshuler, D. Khmel'nitzkii, A.I. Larkin, P.A. Lee, Magnetoresistance and Hall effect in a disordered two-dimensional electron gas, Physical Review B, 22 (1980) 5142-5153.

[51] G. Czack, G. Kirschstein, V. Haase, W.D. Fleischmann, D. Gras, Mo Molybdenum: Physical Properties, Part 2. Electrochemistry, Springer Berlin Heidelberg, 2013.

[52] R. Koepke, G. Bergmann, The upper critical magnetic field Bc2(T) of amorphous molybdenum films, Solid State Communications, 19 (1976) 435-437.

[53] D. Gall, Electron mean free path in elemental metals, Journal of Applied Physics, 119 (2016) 085101.

[54] A. Möbius, C. Frenzel, R. Thielsch, R. Rosenbaum, C.J. Adkins, M. Schreiber, H.D. Bauer, R. Grötzschel, V. Hoffmann, T. Krieg, N. Matz, H. Vinzelberg, M. Witcomb, Metal-insulator transition in amorphous $Si_{1-X}Ni_X$: Evidence for Mott's minimum metallic conductivity, Physical Review B, 60 (1999) 14209-14223.

[55] A. Möbius, Comment on ``Critical behavior of the zero-temperature conductivity in compensated silicon, Si:(P,B)'', Physical Review B, 40 (1989) 4194-4195.

[56] A. Mani, A. Bharathi, Y. Hariharan, Pressure-induced insulator-metal transition of localized states in $FeSi_{1-X}Ge_X$, Physical Review B, 63 (2001) 115103.